# Using Computer Vision for Skin Disease Diagnosis in Bangladesh: Enhancing Interpretability and Transparency in Deep Learning Models for Skin Cancer Classification

Rafiul Islam[a, *], Jihad Khan Dipu[b] and Mehedi Hasan Tusar[c]

[a]Department of Computer Science and Engineering, Daffodil International University, Savar, Bangladesh
[b]Department of Computer Science and Engineering, Daffodil International University, Savar, Bangladesh
[c]Department of Computer Science and Engineering, Daffodil International University, Savar, Bangladesh

***Abstract:*** With over 2 million new cases identified each year, skin cancer is the most prevalent kind of cancer worldwide. Skin cancer is the second most prevalent cancer in Bangladesh, just after breast cancer. To improve patient outcomes, skin cancer must be detected and treated early. The availability of dermatologists and other medical professionals who can identify and cure skin cancer is limited in Bangladesh. Because of this, many skin cancer cases are not discovered until they are well advanced. Images of skin cancer may be successfully classified using deep learning algorithms, according to research. These models, however, often lack interpretability, which may make it challenging to comprehend why they arrive at certain conclusions. The application of deep learning models to enhance skin cancer detection and therapy may be challenging due to this lack of interpretability. In this article, we provide a technique for improving the interpretability of deep learning models for the categorization of skin cancer in the setting of Bangladesh. The characteristics that are crucial for the model's judgments are visualized using a mix of saliency maps and attention maps in our technique. On a collection of photos of skin cancer from Bangladesh, we test our methodology. Our findings demonstrate that our approach may enhance the interpretability of skin cancer categorization deep learning models without materially reducing their accuracy. It also indicate that using our strategy may make deep learning models for identifying skin cancers more interpretable. We may better grasp the reasoning behind the model's judgments by examining the saliency and attention maps. This may be beneficial for medical professionals that use deep learning models to identify and treat skin cancer. Improved skin cancer detection and treatment in Bangladesh may be achieved by using our technique for making deep learning models more interpretable for skin cancer categorization. Any deep learning model for classifying skin cancer may be utilized with our technique, which is straightforward to construct. The patient's age and medical background will be added to the photograph in the future in order to enhance our process. We also want to test our approach using a bigger sample of skin cancer photos. By making deep learning models easier to read, the suggested strategy may aid in improving the detection and treatment of skin cancer in Bangladesh. This might result in early skin cancer identification and treatment, which would benefit patients.

# 1 INTRODUCTION

## *1.1. Motivation*

An extreme hazard to human life is cancer. It could sometimes result in a human being dying for sure. The human body is capable of harboring a variety of malignancies, and skin cancer is one of the deadliest and fastest-growing tumors. It is brought on by a number of things, including drinking alcohol, smoking, having allergies, becoming sick, having viruses, exercising, changing environments, and being exposed to ultraviolet (UV) radiation. UV radiation from the sun has the potential to completely destroy the DNA inside skin cells (Mridha et al., 2023). In addition, skin cancer may also be brought on by unexpected bodily swellings.

The four most common forms of skin cancer are melanoma, basal cell carcinoma, squamous cell carcinoma, and actinic keratoses. According to the WHO, skin cancer will account for one out of every three cancer diagnoses. In the United States, Canada, and Australia, the number of persons receiving a skin cancer diagnosis has been rising at a mostly steady pace over the last several decades (Thomas et al., 2021).

In the United States, it is anticipated that 5.4 million skin cancer cases would be identified annually. Rapid and efficient clinical tests are becoming more and more in demand. A serious kind of cancer known as malignant melanoma originates from melanocytes found in the skin's epidermis. It's possible that this kind of cancer has already spread quickly and is more difficult to cure. Therefore, early skin cancer detection may result in diagnosis and treatment, enhancing survival rates. Various computer-aided diagnostic (CAD) methods have been suggested during the last few decades to detect skin cancer. In order to diagnose cancer, a variety of variables including form, size, color, and texture are extracted using traditional computer vision techniques as a classifier. Artificial intelligence (AI) has now a capability to deal with these issues (Yuan et al., 2021). The most reputable deep-learning architectures, including recurrent neural networks (RNN), deep neural networks (DNN), and convolutional neural networks (CNN), are used in the medical industry to identify cancer cells. These models are also effective in categorizing skin cancer. Additionally, CNN, a DNN in particular, has already produced outstanding success in this area. The most widely used machine learning method for feature learning and object categorization is the CNN model. In order to increase the accuracy of the outcomes, transfer learning is also used in these disciplines to big data sets.

## 1.2. Contribution

The main contribution of our paper is described as follows:

- We suggest a deep convolutional neural network (DCNN) model that can more reliably classify skin cancer even in individuals with early-stage disease.
- The result of our proposed DCNN model achieves much higher accuracy than other existing deep learning (DL) models with a large dataset.
- The required execution time of our proposed DCNN model is very much lower than existing transfer learning models like AlexNet, ResNet, VGG-16, DenseNet, and MobileNet to execute the output results.

1.3. Organization

The remainder of this essay is illustrated as follows: The literature review is presented in Section 2. The difficulties in detecting skin cancer are discussed in Section 3. Section 4 details the materials and procedures. In Section 5, we explain transfer learning using our suggested DCNN model. The training and performance of our suggested DCNN model are shown in Section 6. The results and discussion are presented in Section 7, and the conclusion and suggested next steps are outlined in Section 8.

## 2 LITERATURE REVIEW

One kind of skin cancer that causes malignant tumors on the skin is melanoma. Dermato- logical pictures are used to find skin cancer. Skin cancer is detected with excellent identification accuracy using machine learning based on high-performance images (Srividhya, Sujatha, Ponmagal, Durgadevi, Madhesh-waran, et al., 2020). However, by extracting additional characteristics, the model's accuracy may be improved but its sensitivity is more skewed. The author (Hoshyar, Al-Jumaily, & Hoshyar, 2014) put out a technique that makes use of image processing processes to improve the detection accuracy of skin cancer. They were unable to describe a particular model that is effective in detecting cancer, however. In a different research, the author suggested an architecture-driven model for skin cancer diagnosis that made use of a DL algorithm. The model can anticipate the outcome just as soon since DL based on model-driven architecture can be developed so quickly. It produced superior skin cancer detection results (Kadampur & Al Riyaee, 2020). To advance the medical industry, the technique needs to interface with medical imaging in real time.

The author (Hasan, Barman, Islam, & Reza, 2019) suggested employing CNN for the diagnosis of skin cancer, with the feature being derived from dermoscopic pictures using feature extraction methods. In the testing phase, they had an accuracy of detection of 89.5%. However, improvement was required in the area of detection accuracy. A flaw in the study was that there was overfitting between the testing and training phases (Kourou et al., 2021). To identify and categorize skin cancer, the author of (Li & Shen, 2018) suggested a lesion indexing network (LIN) based on DL. By extracting additional features, they used DL-based LIN to get good results. To further improve results, segmentation performance has to be improved.

The author of (Tschandl et al., 2019) employed CNN to identify pigmented melanocytic lesions as skin cancer from dermoscopic pictures. Screening for non-melanocytic and non-pigmented skin cancer proved challenging, nevertheless. Additionally, its detection accuracy decreased. In (Saba, Khan, Rehman, & Marie-Sainte, 2019), the author proposed a DCNN that includes three steps that work incredibly well to detect skin lesions. The first step uses a color transformation to improve contrast; the second step uses a CNN approach to extract lesion boundaries; and the third step uses transfer learning to extract deep features. Although the strategy produced positive results for certain datasets, the outcomes may vary for other datasets. In (Jafari et al., 2016), the author suggested a CNN-based model to identify melanoma skin cancer. They employed pre-and post-processing of the picture for the enhancement before to and after segmentation, respectively. By merging local and global contextual data, the model generated lesion regions. It achieved a decent classification and prediction performance. The findings may be more valuable if the execution time were disclosed, which is not the case.

Without preprocessing procedures or feature manual selection, Le et al. suggested the more accurate ResNet50 transfer learning model (Le, Le, Ngo, & Ngo, 2020). The findings might be improved with a higher classification rate of skin lesions photos by doing the preprocessing stages, even if precision, recall, and F1 score are not good enough.

## 3  CHALLENGES OF SKIN CANCER DETECTION

Due to the variety of picture kinds and sources, skin cancer detection might be challenging in certain cases. Skin cancer identification is complicated and difficult due to the range in appearance of human skin tone. The most noticeable visual aspects of skin lesions photographs are detailed below, along with a graphic illustration of these difficulties in Fig. 1:

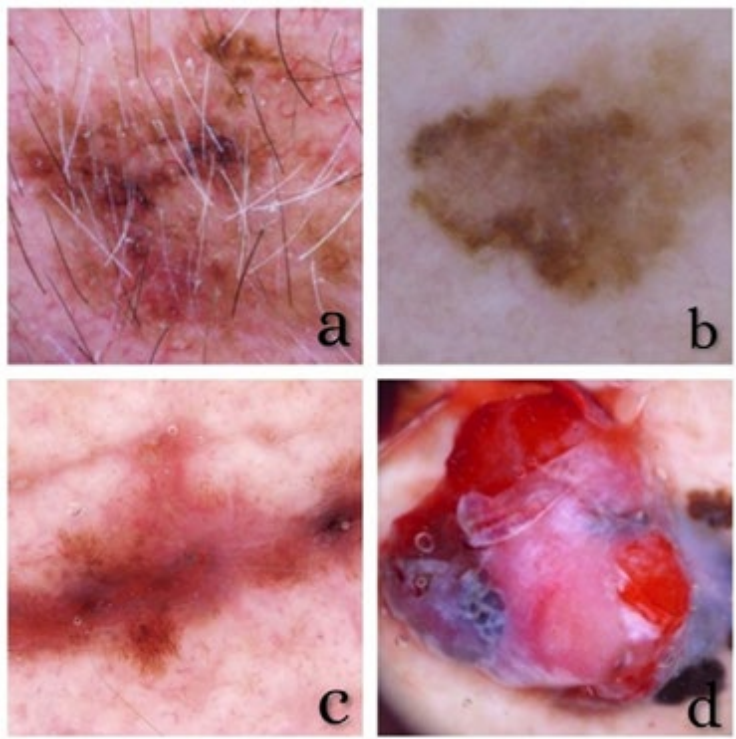

**Fig. 1.** Challenges of skin lesions detection: a: hair artifacts, b: low contrast, c: irregular boundaries, d: color illumination.. (For interpretation of the references to color in this figure legend, the reader is referred to the web version of this article.)

1. The pictures of skin cancer come in a variety of sizes and forms, making it difficult to identify the disease accurately. According to this viewpoint, pre-processing is necessary for precise analysis.

2. In order to get a satisfactory outcome, a few inefficient signals that were not originally a component of the picture must be sacrificed. Therefore, all of this noise and artifacting should be eliminated during the pre-processing processes.

3. Low contrast from nearby tissues might sometimes provide additional challenges and hinder accurate analysis of skin cancer.

4. Color lighting presents additional challenges because of its effects on color texture, light beams, and reflections.

5. Although certain moles on the human body may never grow into cancer cells, they make it more difficult to effectively identify skin cancer from carcinogenic photos.

6. Another issue with detecting skin cancer is the present bias, which alters the performance of the models to provide a better result.

## 4    Proposed Methodology

In this part, we outline the subsequent phases and our methodology's flowchart, which is seen in Fig. 2. When classifying benign from malignant lesions, DCNN and transfer learning models are utilized. The DCNN model we suggest uses several extra layers. On the same dataset, we also compare the performance of several transfer learning models such as AlexNet, ResNet, VGG-16, DenseNet, and MobileNet.

### 4.1. Dataset

Simply said, our dataset is a collection of several skin cancer picture kinds. It needs a lot of data that produces strong results to use DL techniques. The collection of photographs of skin cancer, however, is much more important. Applying DL algorithms in the absence of training data is also one of the primary issues. We used the publicly available dataset known as HAM10000, which contains 10015 dermoscopy images taken from patients in Australia and Austria, to get around these issues. In this dataset, there are 6705 benign lesions, 1113 malignant lesions, and 2197 undetermined lesions. One or more of the following methods—pathology, master agreement, or confocal microscopy—confirmed the ground truth for this dataset. The HAM10000 data that we used in this study were meticulously constructed from melanocytic lesions that were biopsy-proven and classified as benign or malignant (Elyan et al., 2022).

### 4.2. Data preparation

In data preparation, we transform raw data into a more suitable form to create an effective model. Our dataset contains sets of images comparing to similar lesions but from various perspectives or, various images of the same lesions on the same person. Some images were removed from the test and validation sets that are visually blurry and far-away but were still used in training (Yan et al., 2023). There are some factors in data preparation such as data cleaning, feature selection, data transforms, feature engineering, dimensionality reduction, etc that finalize the dataset for training.

Page **5** of 18

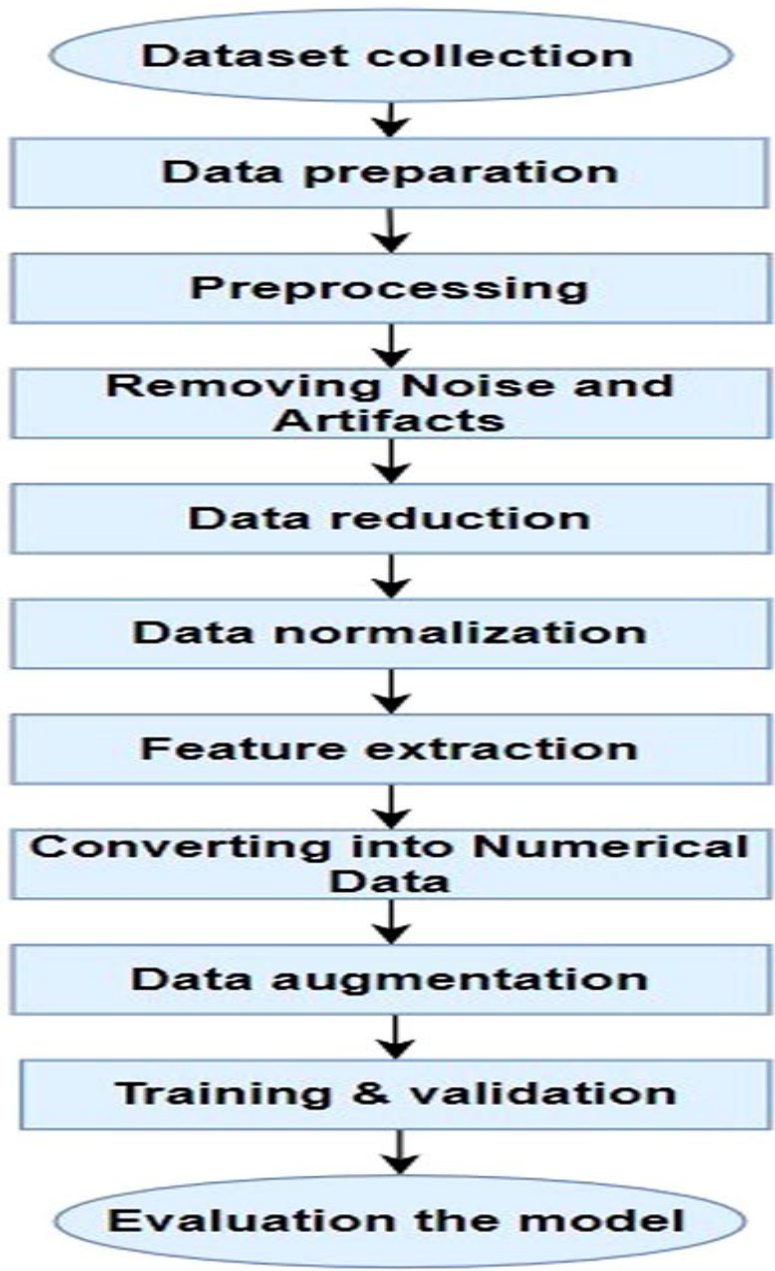

**Fig. 2.** The following steps of our proposed methodology.

*4.3. Preprocessing*

The first and most important stage in preparing the raw data and making it suitable with a machine learning model is data preparation. Preprocessing our dataset's primary goal is to improve the original medical pictures by eliminating air bubbles, noise, and artifacts that are brought on by gel that was applied before to image capture. To achieve a high classification rate in the research we propose, noise and artifacts from the photos were eliminated. In addition, we used data reduction, normalization, feature extraction, and conversion of label string data into numerical data.

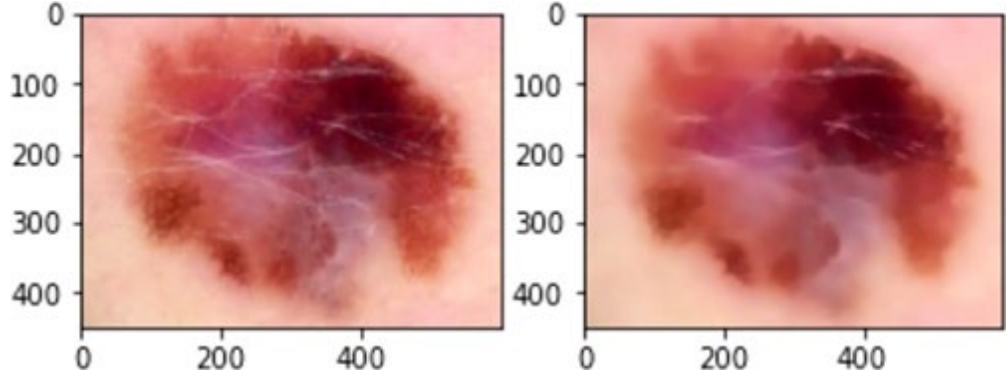

**Fig. 3.** Removing noise and artifacts from the input image.

### 4.1. Reflective noise and artifacts removal

To eliminate even a little amount of noise, there are several kinds of picture smoothing methods available, such as median blurring and Gaussian blurring. A basic thresholding method is employed to remove noise and artifact reflection from the photos. The elimination of noise from the original input picture is shown in Fig. 3.

According to the following criterion, each single pixel (x, y) may be identified and categorized as a reflection artifact:

$$\{I(x, y > T_{R1})\} \text{ and } \{(I(x, y) - I_{avg}(x, y)) > T_{R2}\} \tag{1}$$

where $I$ denotes the image with single-pixel $(x, y)$, $I_{avg}(x, y)$ denotes the average intensity of the pixel's neighborhood which is computed by the local mean filter with dimensions i.e., 12 × 12 and the threshold values which are obtained by experimental i.e., $T_{R1} = 0.87$ and $T_{R2} = 0.096$, respectively.

When the locations of the artifacts are detected, an imprinting activity is performed in a similar manner to the data of the location of the objects' identified pixels, minimizing the influence of the artifacts. There may be an other way to achieve it. In order to do this, imagine a pixel with noise.

$$T_p = T_{p0} + n \tag{2}$$

where $T_{p0}$ = true value of pixel and n= the noise in that pixel. We get
$T_p = T_{p0}$ since the mean of noise and artifacts are zero.

### 4.2. Data reduction

Data reduction is the process of putting fewer photos from the original amount of data in an imaging dataset. Due to the large number of photos with noise and artifacts, some blurry, some with poor contrast, some with moles next to the wounds, and others with color lighting, it is very difficult to get the best classification rate from the whole dataset. We manually extracted several pictures from our proposed DCNN model that have these characteristics. We took into account 6136 photos of benign lesions and 979 images of malignant lesions after subtracting the images.

*4.3. Normalization of data*

The process of designing a database that eliminates data redundancy, data uprightness, and undesirable qualities including insertion, update, and deletion anomalies is known as data normalization. Several normalizing methods are now in use, including min-max normalization, z-score normalization, and decimal scaling normalization.

*4.1. Feature extraction*

In order to divide images into more manageable groups for further processing, the method of feature extraction is essential. In our study, we extract a significant number of characteristics that aid in identifying and recognizing the pattern of several datasets. Additionally, it chooses and mixes variables to extract features that use less resources while preserving all of the original data's information.

*4.2. Working with numerical data*

The most common sorts of data that will be dealt with in machine learning algorithms for imputation are numerical values. We have a method for obtaining numerical values for each characteristic that have different scales. For easier processing in the training and model supports with a broad variety of DL network typologies, the data must also be normalized and standardized. The LabelEncoder function from the Python standard library was used in our experiment to translate the two labels, benign and malignant, into the numbers 0 and 1.

*4.3. Data augmentation*

The process of adding slightly changed copies of current data without actually gathering new data from existing training data is known as data augmentation. The training dataset size may be intentionally increased via data warping or oversampling, which also helps the model avoid over-fitting by preventing it from the source of the issue. We used principal component analysis to augment our data using various augmentation settings, including rotation, random cropping, mirroring, and color-shifting, in order to combat this over-fitting. Table 1 provides examples of the settings we employed for our dataset's data augmentation.

## 5     TRANSFER LEARNING WITH OUR PROPOSED DCNN MODEL

A machine learning approach called transfer learning allows the pre-trained model to be used to many, related tasks (Singh et al., 2020). A new DNN model must be trained on a huge number of photos (Madanu et al., 2022). Unfortunately, there aren't any datasets out there with a ton of labeled pictures of skin lesions. Using all of the neural network's available parameters to train a big medical dataset like ImageNet is a difficult undertaking. However, we used a large medical dataset to train AlexNet, ResNet, VGG16, DenseNet, MobileNet, and our suggested DCNN model from our research.

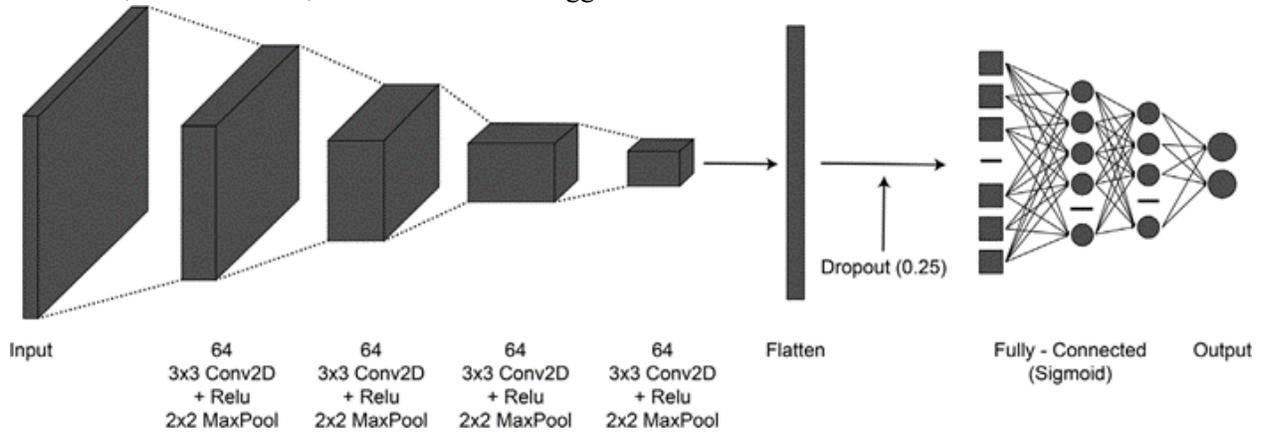

**Fig. 4.** Model architecture of our proposed DCNN model.

Table 1

Using data augmentation parameters and values.

| Data augmentation parameter | Parameter value | Action |
| --- | --- | --- |
| Rotation_range | 10 | Input data generates with the rotation from -10 to 10 |
| Height_shift_range | 0.2 | Image is randomly shifted vertical direction by 0.2 |
| Width_shift_range | 0.2 | mage is randomly shifted in horizontal direction by 0.2 |
| Shear_range | 0.2 | Stretch the image angle slantly in degrees by a factor of 0.2 |
| Zoom_range | 0.2 | Zoom in or out 0.2 from the center |
| Channel_shift_range | 10 | Randomly shifts channel values to variate the color |
| Horizontal_flip | True | Flips the image horizontal direction randomly |

| Fill_mode | Nearest | Closest pixel value is chosen to fill the empty values |

In ImageNet, the classification value for the final layer is 1000, compared to 10 in AlexNet and 2 in the suggested DCNN model we utilized to distinguish between malignant and benign skin lesions. So, rather of using a softmax, we employed a sigmoid layer. The back-propagation function is used for fine-tuning to get the updated weights in a classification method that is more effective. In our suggested DCNN model, Adam optimizer is employed as a gradient descent approach. Finally, an augmentation process is used to get beyond the dataset's labeled image constraints. The architecture of our suggested DCNN model is shown in Fig. 4.

## 6    TRAINING AND PERFORMANCE

Our suggested DCNN model is tuned using a number of parameters. The main ones we employed in our experiment, together with the values we picked for each, are listed below. This setup is contrasted with a variety of previously tested deep neural network models and topologies.

- The pooling layer (MaxPooling2D): is used to reduce the size of input images for faster training.
- Batch size (128): the number of processed images in every iteration.
- Initial learning rate (0.001): sets up the rate at which the learning procedure will begin.
- Validation frequencies (36): the number of iterations between eval- uations of the validation metrics.
- Optimizer (Adam): Adam is a substitution optimization algorithm
- for stochastic gradient descent in order to minimize the loss
- function for training DL models. In our research, we selected the
- Standard gradient descent algorithm with a momentum of 0.999.
- Loss function (binary cross-entropy): it is used to set up a binary classification problem from the dataset.
- Number of epochs (100): the number of times the dataset is passed to the DNN.

- On the HAM10000 dataset, the simple and progressive models were recently trained and fine-tuned to increase their understanding of the skin disease categorization issues. In order to optimize and assess the suggested DCNN model, we divided the dataset into three sets for training, validation, and testing in our study.

- For a better comparison between transfer learning methods and our suggested DCNN model, the dataset is split into two groups: 70% for training, 20% for validation, and 10% for testing; and 80% for training, 10% for validation, and 10% for testing. The model was implemented on a machine with an Intel Core i5-8250U CPU processor, 8 GB of RAM, and an NVIDIA GeForce MX130 GPU card using a Jupyter notebook and Google Colab (12 GB RAM). In this discipline, accuracy, precision, recall, F-measure, and execution time are used to evaluate each model's performance to that of the current neural network. The confusion matrix, a table of proposed output for evaluating the model performances, is also shown in Fig. 5. The confusion matrix includes the following four key terms:

- True Positives (TP): The occurrence in which we anticipated true and the real yield was likewise true.
- True Negatives (TN): The occurrence in which we anticipated false and the real yield was likewise false.
- False Positives (FP): The occurrence in which we anticipated true and the real yield was likewise false.
- False Negatives (FN): The occurrence in which we anticipated false and the real yield was likewise true.

Precision: It is known as the number of right positive results partitioned by the number of positive outcomes anticipated by the classifier.

$$Precision = TP / TP + FP \qquad (6)$$

Recall: It is the quantity of right sure outcomes divided by the number of every conjugate samples (all samples that ought to have been identified as sure).

$$Recall = TP / TP + FN \qquad (7)$$

F1-score: It is also known as harmonic mean that measures to seek a balance between precision and recall. It takes both false positives and false negatives for the computation and performs well on an imbalanced dataset.

$$F1 - score = 2TP / 2TP + FP + FN \qquad (8)$$

Accuracy: It refer to the ratio of numbers of correct predictions to the total number of input samples.

$$Accuracy = TP + TN / TP + TN + FP + FN \qquad (9)$$

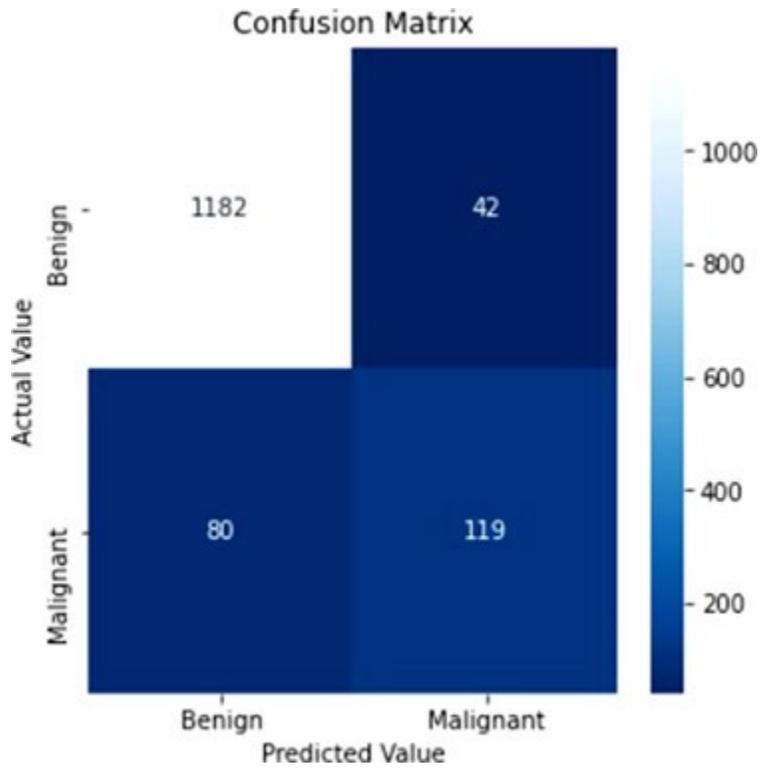

**Fig. 5.** Regarding performance, confusion matrix of the proposed DCNN model

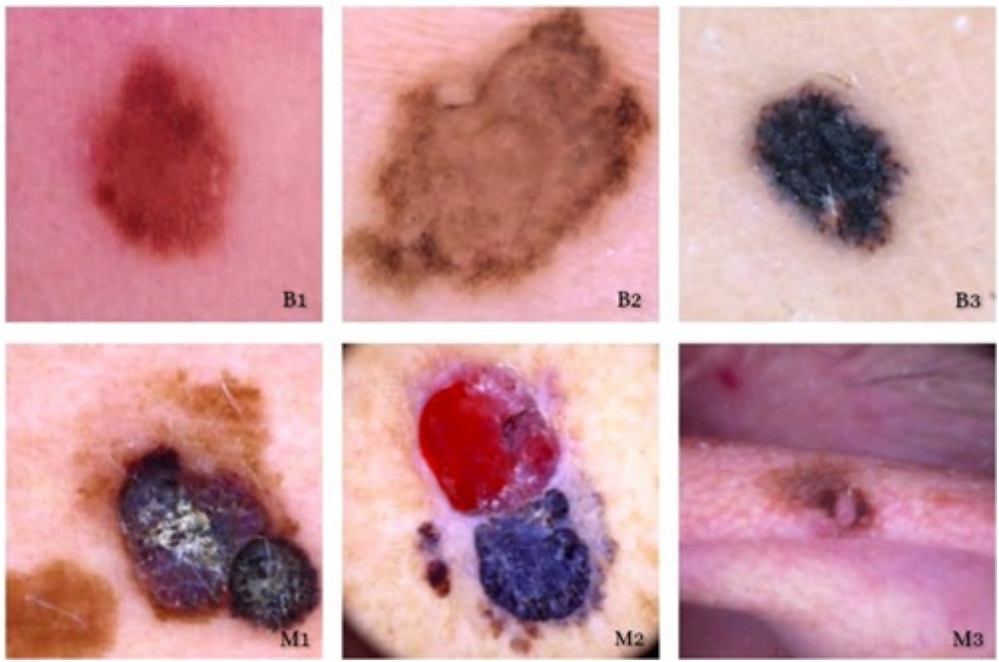

**Fig. 6.** B1, B2, and B3 are benign, and M1, M2, and M3 are malignant skin lesions, respectively.

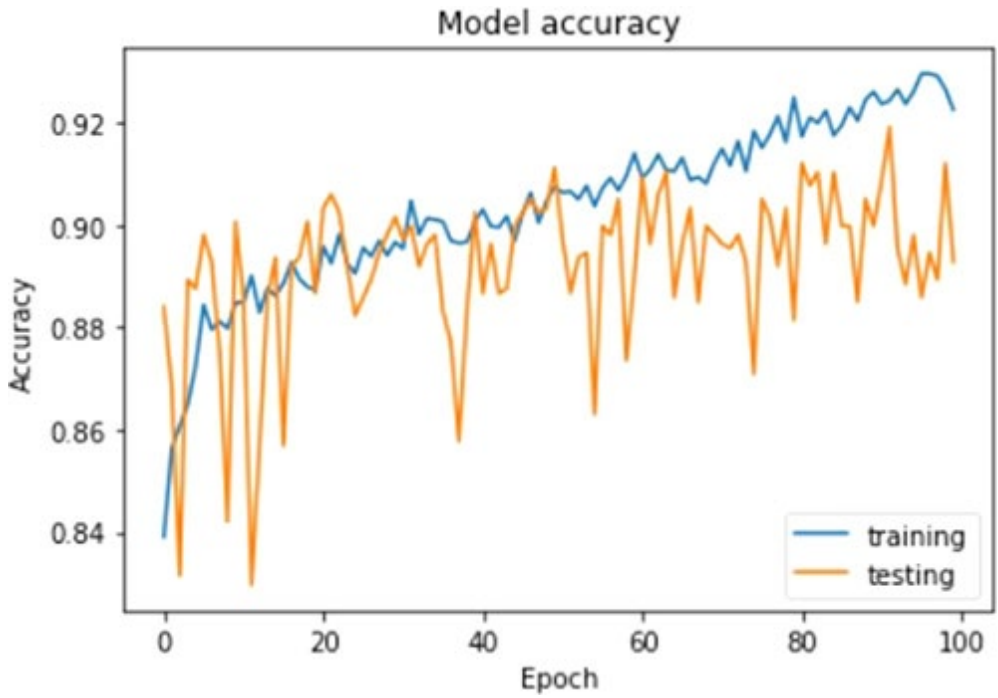

**Fig. 7.** Performance curve based on accuracy per epoch of our proposed DCNN model.

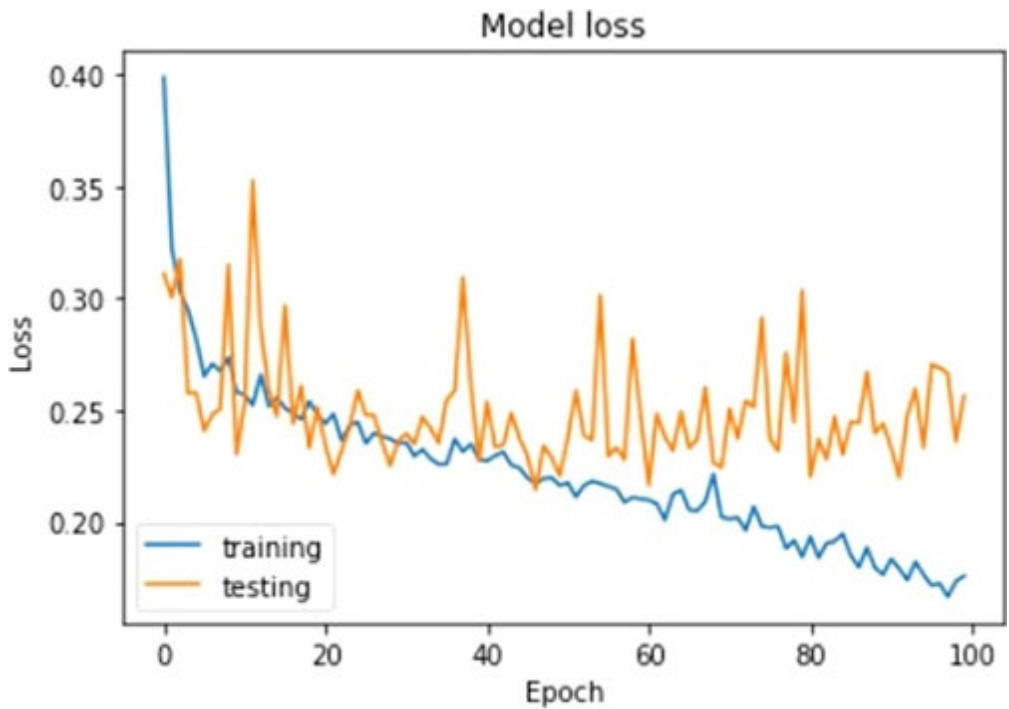

**Fig. 8.** Performance curve based on loss per epoch of our proposed DCNN model.

# 7 RESULT AND DISCUSSION

Our proposed model's primary function is to distinguish between benign and malignant skin lesions from the dataset that was retrieved using DCNN (Zhang et al., 2021). The total number of photographs we acquired from the Internet following the data reduction process is used to determine the findings. In Fig. 6, several benign and cancerous pictures are shown.

In this research, we tested our suggested DCNN model with two sets of training photos: 70% and 80%, and the results reveal that the 80% set of training images has the greatest accuracy. On the same dataset, we also tested a number of other DNN models, including AlexNet, ResNet, VGG-16, DenseNet, and MobileNet, however our suggested DCNN model had the highest classification accuracy. The accuracy and loss between the training and testing phases as determined by our suggested DCNN model are shown in Figs. 7 and 8, respectively.

In Tables 2 and 3, three transfer learning models with two different kinds of split data are compared to the outcomes of our proposed DCNN model.

Based on numerical performance and visual findings, a thorough explanation of the suggested DCNN model is done. With the aid of image processing methods that are discussed in Sections 4.3 to 4.9, we were able to achieve such an excellent classification rate in this model. Our suggested model was also run with 200 epochs, however at that point the model was becoming over-fitted. The area under the curve (AUC) value of 0.847 achieved with 100 epochs after multiple iterations of fine-tuning is shown in Fig. 9. A CAD system that uses the suggested DCNN model may efficiently identify skin lesions at an early stage. Additionally, the early identification of a malignant skin growth may greatly increase the chance of survival, particularly for people who lack access to medical care.

**Table 2**

The performance comparison between our proposed DCNN model and transfer learning models taking 70% of training, 20% of validation, and 10% of testing data.

| Model | Precision | Recall | F1 score | Training Acc | Testing Acc |
|---|---|---|---|---|---|
| AlexNet (Han, Zhong, Cao, & Zhang, 2017) | 96.89 | 90.70 | 93.70 | 92.05 | 88.81 |
| ResNet (Targ, Almeida, & Lyman, 2016) | 84.94 | 97.50 | 90.79 | 92.78 | 85.20 |
| VGG-16 (Guan et al., 2019) | 88.27 | 95.18 | 91.59 | 88.83 | 86.09 |
| DenseNet (Carcagnì et al., 2019) | 91.00 | 91.75 | 91.37 | 91.36 | 85.25 |
| MobileNet (Sinha & El-Sharkawy, 2019) | 84.62 | 94.57 | 89.32 | 92.93 | 82.62 |
| Proposed DCNN model | 94.63 | 93.91 | 94.27 | 92.69 | 90.16 |

**Table 3**

The performance comparison between our proposed DCNN model and transfer learning models taking 80% of training, 10% of validation and 10% of testing data.

| Model | Precision | Recall | F1 score | Training Acc | Testing Acc |
|---|---|---|---|---|---|
| AlexNet (Han, Zhong, Cao, & Zhang, 2017) | 97.88 | 92.01 | 94.85 | 93.82 | 90.86 |
| ResNet (Targ, Almeida, & Lyman, 2016) | 96.00 | 93.63 | 94.80 | 91.25 | 90.93 |
| VGG-16 (Guan et al., 2019) | 1.00 | 86.02 | 92.48 | 86.17 | 86.02 |
| DenseNet (Carcagnì et al., 2019) | 91.42 | 94.83 | 93.09 | 91.64 | 88.33 |
| MobileNet (Sinha & El-Sharkawy, 2019) | 92.08 | 94.15 | 93.10 | 90.31 | 88.26 |
| Proposed DCNN model | 96.57 | 93.66 | 95.09 | 93.16 | 91.43 |

**Table 4**

Comparison of computational time $t$ over the existing models.

| Model | Time per epoch | Total time |
|---|---|---|
| AlexNet (Han, Zhong, Cao, & Zhang, 2017) | 14-16 s | 24+1 min |
| ResNet (Targ, Almeida, & Lyman, 2016) | 13-15 s | 23+1 min |
| VGG-16 (Guan et al., 2019) | 14-15 s | 24+1 min |
| DenseNet (Carcagnì et al., 2019) | 16-18 s | 28+1 min |
| MobileNet (Sinha & El-Sharkawy, 2019) | 11-12 s | 19+1 min |
| Proposed DCNN model | 9-10 s | 16±1 min |

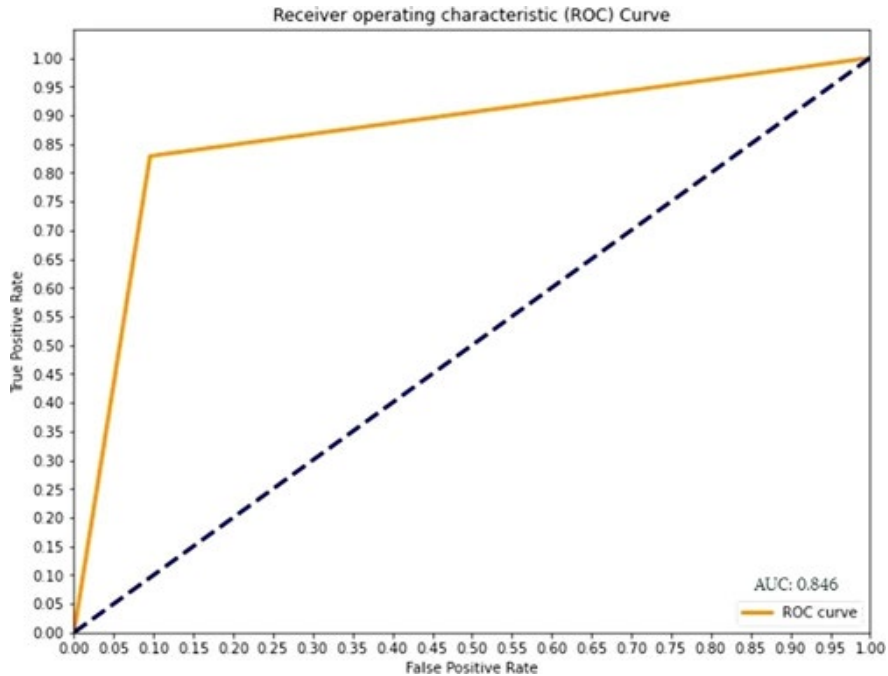

**Fig. 9.** Receiver Operating Characteristic curve of our proposed DCNN model.

## 8    Conclusion and Future Work

In this study, the suggested DCNN model outperforms existing transfer learning models in terms of classification accuracy. The suggested technique has the capacity to categorize benign and malignant skin lesions by substituting a sigmoid for the output activation layer for binary classification. Additionally, the suggested approach was tested on the HAM10000 dataset, where we found that it outperformed previous transfer learning models in terms of training and testing accuracy. Additionally, the model's ability to improve its accuracy was hindered by the unbalanced dataset and the lack of many photos. We balanced the dataset for both levels as a consequence, which increases classification accuracy. On the same dataset, we also trained various transfer learning models, but the outcomes weren't any better than those of our suggested DCNN model. Transfer learning models sometimes performed well, but they required more time to execute every epoch than our suggested DCNN model, as is seen in Table 4. To the best of our knowledge, there is no existing research that we are aware of that classified skin lesions as benign or malignant using the same dataset.

## 9    Acknowledgement